\documentclass[a4paper,11pt]{article}
\usepackage[a4paper, margin=0.7in]{geometry}
\usepackage{jheppub} 
\usepackage{setspace}
\usepackage{amsmath}
\usepackage{graphicx}
\usepackage{relsize}
\usepackage{empheq}
\usepackage{physics}
\title{\boldmath Multi-soft theorems for cosmological correlators: Background wave method for scalars \& gravitons}

\author{Farman Ullah}
\affiliation{International Centre for Theoretical Sciences - TIFR,\\
Bengaluru - 560089, India}

\emailAdd{farman.ullah@icts.res.in}

\abstract{Cosmological soft theorems (or consistency relations) provide a powerful probe for the physics of inflation. These relations rely on minimal assumptions and hold very generally. Consequently, any violation of these relations would rule out a large class of inflationary models. For instance, a violation of the scalar soft theorem (or consistency relation) would rule out all attractor single-field inflation models and instead point toward either multi-field dynamics or a non-attractor phase. In this paper, we derive tree-level multi-soft theorems, at leading order in the soft expansion, for both scalar and tensor correlation functions. Our analysis employs the background-wave method, in which the effect of long-wavelength modes is captured by an appropriate spatial coordinate rescaling. In addition, we systematically incorporate soft-exchange contributions, including tensor exchanges in scalar correlators and scalar exchanges in tensor correlators.}

\begin{document}
\maketitle
\flushbottom
\section{Introduction}
Inflation \cite{Starobinsky:1980te,Guth:1980zm,Linde:1981mu,Starobinsky:1979ty,Mukhanov:1981xt} has proven to be a simple and robust mechanism to generate perturbations in the early universe with a spectrum consistent with observations. However, the exact nature of this phase remains elusive. Although in the simplest models, only a single scalar degree of freedom is active during inflation, inflation could, in principle, be driven by multiple scalar fields. There might even be other exotic kinds of field content active during this phase \cite{Bordin:2016ruc,Lee:2016vti,Baumann:2017jvh}. The cosmological consistency relations or soft theorems \cite{Maldacena:2002vr,Weinberg:2003sw,Creminelli:2004yq,Cheung:2007sv,Creminelli:2012ed,Hinterbichler:2012nm,Creminelli:2013cga,Goldberger:2013rsa,Hinterbichler_2014} are an excellent probe for these kinds of questions because they rely on very few assumptions. For instance, the scalar soft theorem relies just on the fact that there is a single scalar degree of freedom during inflation, and it freezes on super-Hubble scales (non-attractor models are an exception \cite{Kinney:2005vj,Huang:2013oya,Chen:2013eea,Martin:2012pe,Namjoo:2012aa}, see \cite{Mooij:2015yka,Finelli:2017fml,Finelli:2018upr} for modified soft theorems in shift symmetric non-
attractor models). Therefore, a violation of these relations will point towards scenarios with multi-field inflation or a non-attractor phase. The tensor single-soft theorem, on the other hand, is even more robust \cite{Bordin:2016ruc}, i.e, it holds in the presence of multiple scalar fields and tensor perturbations which decay fast enough outside the horizon. A violation of such a relation will imply that during inflation, there was some residual anisotropy that didn't wash out. This might provide a useful window into exotic scenarios such as partially massless fields \cite{Baumann:2017jvh}. \\\\
In this paper, \textit{using the background wave method, we derive tree-level multi-soft theorems for scalars and gravitons} during inflation. The $N=2$ \& $3$ soft theorems we derive capture the full soft limit (leading order in the soft expansion) of the correlation functions. The $N>3$ soft theorems include single soft exchanges; they do not include diagrams with double, triple, or higher soft exchanges. This is also true for previous work on multi-soft theorems \cite{Senatore_2012,Joyce:2014aqa}. In single-field inflationary models where contributions from single soft exchange diagrams dominate, our results are expected to provide an accurate—effectively complete—description of the multi-soft limit. However, when this is not the case, our results capture only a subset of the contributions to the multi-soft limit. It is non-trivial to include all possible soft exchanges, and we leave it for future work. We will consider correlators of the kind $\langle\zeta^N\zeta^M\rangle$ with $N$ of the scalars taken soft and $\langle\gamma^N\zeta^M\rangle$ with $N$ gravitons taken soft. By soft, we mean that the momenta are much smaller compared to the rest. We do not assume any hierarchy among the soft momenta. Multi-soft theorems for scalar correlators have been extensively discussed in the literature \cite{Chen:2006dfn,Senatore_2012,Mirbabayi:2014zpa,Joyce:2014aqa,Alinea:2016glw}. In particular, a multi-soft theorem for scalar correlators was derived in \cite{Joyce:2014aqa} using the $1$PI action method. Here, we derive this relation solely from the background wave method. Unlike \cite{Joyce:2014aqa}, we also include graviton exchange contributions (two or more soft external scalar modes combine into an internal soft graviton mode). Similar contributions for the scalar double-soft theorem were considered in \cite{Alinea:2016glw}. The multi-soft theorem we derive for gravitons is novel and, to the best of our knowledge, has not appeared in the literature so far. This relation (as we will see), unlike the tensor single-soft theorem, is sensitive to adiabaticity (super-Hubble freezing) of both scalar and tensor modes. Therefore, violations of these relations can arise from tensor non-adiabaticity, the presence of multiple scalar fields, or a non-attractor phase during inflation. \\\\
The rest of the paper is organised as follows. In Sec. \ref{scalar:single:soft}, we review the background wave method in the case of the single-soft theorem for scalar correlators. In Sec. \ref{double:soft:gravitons} we derive the double-soft theorem for graviton-scalar correlators (two gravitons taken soft). In Sec. \ref{N:soft:scalars} we derive the $N$-soft theorem for scalar correlators ($N$ scalars taken soft). We urge the reader to read Sec. \ref{N:soft:scalars} before moving on to Sec. \ref{n:soft:gravitons}. In Sec. \ref{n:soft:gravitons} we derive the $N$-soft theorem for graviton-scalar correlators ($N$ gravitons taken soft). We conclude in Sec. \ref{conclusion}.
\paragraph{Notations and conventions}
Throughout this paper, we work in the so-called co-moving gauge where the inflaton remains unperturbed ($\delta\phi(x)=0$), and the spatial metric becomes $h_{ij}=a^{2}(t)e^{2\zeta}(e^{\gamma})_{ij}$. Here, $\zeta$ is the curvature perturbation (scalar degree of freedom) and $\gamma_{ij}$ is the graviton (tensor degree of freedom). We work in a gauge where the graviton is transverse and traceless, i.e., $\gamma_{ii}=0,\hspace{2pt}\partial_i\gamma_{ij}=0$. Conformal time is denoted by $\eta$ with the end of inflation being $\eta\rightarrow 0$. It would be useful for us to define two kinds of differential operators since they will appear frequently in our formulae: 
\begin{align}
\delta_{\zeta}Z&=\left(-3(M-1)-k^i_{a}\partial_{k^i_a}\right)Z \\
\delta^{\gamma}_{ij}Z&=-\frac{1}{2}k^i_{a}\partial_{k^j_a}Z
\end{align}
where $Z$ is the stripped correlator of $M$ hard modes in Fourier space. The sum over $i,j$ runs from 1 to 3 (sum over components of each momentum vector), and for $a$, it runs from 1 to $M$.
\section{Scalar single-soft theorem: a review}\label{scalar:single:soft}
To derive our soft theorems, we employ the so-called background-wave method. The central idea behind this method is that once the (scalar ($\zeta$)/tensor ($\gamma$) as defined in the co-moving gauge) modes leave the horizon, they stop evolving in time. This is a theorem in single-field inflation \cite{Starobinsky:1985ibc,Maldacena:2002vr,Wands:2000dp,Weinberg:2003sw,Lyth:2003im,Pimentel:2012tw,Assassi:2012et,Senatore:2012ya}. Therefore, they can be absorbed in spatial coordinate rescaling, giving rise to soft theorems. We first review the single soft theorem for scalars. As a review, we illustrate the background wave method for the simplest case: a single soft theorem for scalars. At any given time, we can write $\zeta=\zeta_L+\zeta_S$ (long+short modes). The long modes exit the horizon much earlier than the short modes. By the time short modes are crossing the horizon, the long modes are way outside the horizon. They are constant in time and also have weak spatial dependence compared to the Hubble scale; therefore, they can be treated as constant to leading order. The spatial part of the line element of the background metric, therefore, can be written as,
\begin{align}
    a^{2}(t)e^{2\zeta_L}dx^{i}dx^{j}=a^{2}(t)d\tilde{x}^{i}d\tilde{x}^{j}\,,
\end{align}
where, $\tilde{x}^{i}=e^{\zeta_L}x^{i}$. To linear order we have $\tilde{x}^{i}=(1+\zeta_L)x^{i}$. We start by considering the following position space correlation function,
\begin{align}
\langle\zeta_{L}(\vec{x}_1)\zeta(\vec{x}_2)\dots\zeta(\vec{x}_N)\rangle
\end{align}
This correlation function can be computed first by fixing the long mode and then averaging over it,
\begin{align}
\langle\zeta_{L}(\vec{x}_1)\zeta(\vec{x}_2)\dots\zeta(\vec{x}_N)\rangle=\langle\langle\zeta(\vec{x}_2)\dots\zeta(\vec{x}_N)\rangle_{\zeta_{L}}\zeta_{L}\rangle
\end{align}
This can be stated more clearly in terms of Probability Distribution Functionals (PDFs). Given a PDF for the long and short modes, $P[\zeta_L,\zeta_S]$. The correlation function above is then computed as,
\begin{align}
    \langle\zeta_{L}(\vec{x}_1)\zeta_S(\vec{x}_2)\dots\zeta_S(\vec{x}_N)\rangle
    &=\int D\zeta_L D\zeta_S \hspace{2pt} \zeta_L(\vec{x}_1)\zeta_S(\vec{x}_2)\dots\zeta_S(\vec{x}_N)P[\zeta_L,\zeta_S]\nonumber \\
    &=\int D\zeta_L D\zeta_S\hspace{2pt} \zeta_L(\vec{x}_1)\zeta_S(\vec{x}_2)\dots\zeta_S(\vec{x}_N)P[\zeta_S|\zeta_L]P[\zeta_L] \nonumber \\
    &=\int D\zeta_L \zeta_L(\vec{x}_1)P[\zeta_L]\left(D\zeta_S \hspace{2pt} \zeta_S(\vec{x}_2)\dots\zeta_S(\vec{x}_N)P[\zeta_S|\zeta_L]\right)\nonumber \\
    &=\int D\zeta_L \zeta_L(\vec{x}_1)P[\zeta_L]\langle \zeta_S(\vec{x}_2)\dots\zeta_S(\vec{x}_N)\rangle_{\zeta_L} \nonumber \\
    &=\langle\zeta_L(\vec{x}_1)\langle \zeta_S(\vec{x}_2)\dots\zeta_S(\vec{x}_N)\rangle_{\zeta_L}\rangle
\end{align}
where $P[\zeta_S|\zeta_L]$ is a conditional PDF which computes the probability of a short mode given a particular realisation of the long mode. The correlator $\langle\zeta(\vec{x}_2)\dots\zeta(\vec{x}_N)\rangle_{\zeta_{L}}$ is computed at late times ($\eta\rightarrow0$ or $t\rightarrow \infty$). In this limit, all modes are outside the horizon, i.e., wavelength $> H^{-1}$; however, the dominant contribution is picked near the horizon crossing of short modes. Here, the assumption is that the short modes are not of very different wavelengths, so they cross the horizon around the same time. At this time, the modes in $\zeta_L$ are outside the horizon and are therefore frozen. One can then absorb the long mode by spatial coordinate rescaling (as discussed above) and get,
\begin{align}
    \langle\zeta(\vec{x}_2)\dots\zeta(\vec{x}_N)\rangle_{\zeta_{L}}=\langle\zeta(\vec{\tilde{x}}_2)\dots\zeta(\vec{\tilde{x}}_N)\rangle
\end{align}
Now we Taylor expand to linear order in $\zeta$,
\begin{align}\label{single:soft:taylor}
\langle\zeta(\vec{\tilde{x}}_2\dots\zeta(\vec{\tilde{x}}_N)\rangle=\langle\zeta(\vec{x}_2)\dots\zeta(\vec{x}_N)\rangle+\zeta_L(\vec{x}_{+})x_a^{i}\partial_{x_a^{i}}\langle\zeta(\vec{x}_2)\dots\zeta(\vec{x}_N)\rangle
\end{align}
where $a$ runs from 2 to $N$. The long mode is evaluated at the mid point for convenience (it is very slowly varying in space), $\vec{x}_{+}=\frac{\sum_{a}\vec{x}_a}{N-1}$. We proceed by multiplying $\zeta_L(\vec{x}_1)$ and taking an average. The first term gives a one-point function of the long mode, which is zero by definition. Therefore, we get,
\begin{align}
\langle \zeta_{L}(\vec{x}_1)\langle\zeta(\vec{\tilde{x}}_2)\dots\zeta(\vec{\tilde{x}}_N)\rangle\rangle&=\langle\zeta_{L}(\vec{x}_1)\zeta(\vec{x}_2)\dots\zeta(\vec{x}_N)\rangle \nonumber \\
& =\langle\zeta_{L}(\vec{x}_1)\zeta_{L}(\vec{x}_{+})\rangle x_a^{i}\partial_{x_a^{i}}\langle\zeta(\vec{x}_2)\dots\zeta(\vec{x}_N)\rangle
\end{align}
We write both sides in terms of their Fourier transforms,
\begin{align}\label{deltafunction}
 &\int \prod_{i=1}^{N}d^{3}\vec{k}_i \langle\zeta(\vec{k}_1)\dots\zeta(\vec{k_N})\rangle'\delta^{(3)}\left(\sum_{i=1}^{N}\vec{k}_i\right)e^{i\sum\vec{k_a}.\vec{x}_a}\nonumber \\
& =\int d^{3}\vec{k}_{+}\prod_{i=1}^{N}d^{3}\vec{k}_ie^{i\left[\sum_{a=1}^{N}\vec{k}_a.\vec{x}_a+\vec{k}_{+}.\vec{x}_{+}\right]}\langle\zeta(\vec{k_1})\zeta(-\vec{k}_1)\rangle'\delta_{\zeta}\langle\zeta(\vec{k}_2)\dots\zeta(\vec{k}_N)\rangle'
\nonumber \\& 
\hspace{10pt}\times \delta^{(3)}(\sum_{i=2}^{N}\vec{k}_i)\delta^{(3)}(\vec{k}_1+\vec{k}_{+})
\end{align}
where, $\delta_{\zeta}=-3(N-2)-k_{ai}\partial_{k_{ai}}$. Primes $\langle\rangle'$ denote that the momentum-conserving delta function has been factored out. Now we make a few modifications to the RHS. First, we evaluate the $\vec{k}_2$ integral. After this we substitute $\vec{k}_{+}=\sum_{i=2}^{N}\vec{k}_i$. This modifies the RHS to the following:
\begin{align}
    =\int \prod_{i=1}^{N}d^{3}\vec{k}_ie^{iA}\langle\zeta(\vec{k_1})\zeta(-\vec{k}_1)\rangle'\delta_{\zeta}\langle\zeta(\vec{k}_2)\dots\zeta(\vec{k}_N)\rangle' \nonumber \\\times \delta^{(3)}(\sum_{i=1}^{N}\vec{k}_i)
\end{align}
where, \begin{align}A=\vec{k}_1.\vec{x}_1-(\vec{k}_3+\vec{k}_4+\dots+\vec{k}_{N}).\vec{x}_2+\vec{k}_3.\vec{x}_3+\dots+\vec{k}_N.\vec{x}_N\nonumber \\ 
+(\vec{k}_2+\vec{k}_3+\dots+\vec{k}_N)\frac{(\vec{x}_2+\vec{x}_3+\dots+\vec{x}_N)}{N-1}
\end{align}
Notice, we have a delta function $\delta^{(3)}(\sum_{i=1}^{N}\vec{k}_i)$, this can be used to simplify $A$ further,
\begin{align}
    A=\vec{k}_1.\vec{x}_1+\left(\vec{k}_2+\vec{k}_1-\frac{\vec{k}_1}{N-1}\right).\vec{x}_2+\left(\vec{k}_3-\frac{\vec{k}_1}{N-1}\right).\vec{x}_3\nonumber \\
    +\left(\vec{k}_4-\frac{\vec{k}_1}{N-1}\right).\vec{x}_4+\dots+\left(\vec{k}_N-\frac{\vec{k}_1}{N-1}\right).\vec{x}_N\nonumber\\
    \approx \sum_{i=1}^N\vec{k}_i.\vec{x}_i
\end{align}
where we have used the fact that $\vec{k}_1$ is much smaller than all other momenta. Now, we can write the soft theorem in terms of stripped correlators alone as,
\begin{equation}\boxed{\begin{aligned}
\lim_{\vec{k}_1\rightarrow 0}\langle\zeta(\vec{k}_1\dots\zeta(\vec{k_N})\rangle'=\langle\zeta(\vec{k_1})\zeta(-\vec{k}_1)\rangle'\delta_{\zeta}\langle\zeta(\vec{k}_2)\dots\zeta(\vec{k}_N)\rangle'
\end{aligned}}\end{equation}
The derivation for the single soft theorem for the graviton is very similar, and we do not discuss that here.
\section{Double-soft: gravitons}\label{double:soft:gravitons}
The double soft theorem for scalars was derived in \cite{Joyce:2014aqa,Mirbabayi:2014zpa,Alinea:2016glw}. We do not review their calculation here and instead move on to compute the soft theorem for correlators where two gravitons are taken to be soft. We start again by considering the spatial part of the line element in the co-moving gauge,
\begin{align}
    a^{2}(t)e^{2\zeta_L}\left(e^{\gamma^L}\right)_{ij} dx^{i}dx^{j}
\end{align}
where $\gamma^L$ and $\zeta_L$ are the long part of the tensor and scalar perturbations. These are constant (up to small corrections), and they can be absorbed in an anisotropic rescaling of the spatial coordinates. Since we will be deriving the soft theorem with two soft modes, we expand the exponential $e^{\gamma^{L}}$ up to quadratic order in $\gamma^L$. We also retain the scalar mode $e^{2\zeta}$ and expand up to linear order. This will give a contribution when two soft graviton modes combine into an internal soft scalar mode. The coordinate transformation which absorbs these fluctuations reads as,
\begin{align}
    \tilde{x}^{i}=x^{i}+\zeta_Lx^{i}+x^{j}\left(\frac{1}{2}{\gamma^L}_{ij}+\frac{1}{8}\gamma^L_{ik}\gamma^L_{kj}\right)
\end{align}
A similar procedure was followed in \cite{Alinea:2016glw}, where instead they computed a soft theorem for the scalar correlation functions while keeping track of the long tensor mode as well. In the new coordinates, the metric in (1) becomes,
\begin{align}
    a^{2}(t)\delta_{ij}d\tilde{x}^{i}d\tilde{x}^{j}
\end{align}
Therefore, the presence of long modes has been absorbed in the background solution. Now, consider the correlation function $\langle\gamma^L_{ij}(\vec{x}_1)\gamma^L_{kl}(\vec{x}_2)\mathcal{O}\rangle$,
where $\mathcal{O}$ is the product of $N$ scalar operators $\prod_{i=1}^{N}\zeta(\vec{y}_i)$. We now proceed to compute this correlator as follows: We first compute $\langle\mathcal{O}\rangle$ in the presence of the long modes (keeping long modes fixed), we call this $\langle\delta\mathcal{O}\rangle$, and then average over the two long modes. Now comes the key point: since the presence of long modes is equivalent to a coordinate rescaling, therefore $\langle\delta\mathcal{O}(\vec{y}_i,\dots,\vec{y}_N)\rangle=\langle\mathcal{O}(\vec{\tilde{y}}_1,\dots,\vec{\tilde{y}}_N)\rangle$. Taylor expanding this gives,
\begin{align}
   \langle\mathcal{O}(\vec{\tilde{y}}_1,\dots,\vec{\tilde{y}}_N)\rangle= \langle\mathcal{O}(\vec{y}_1,\dots,\vec{y}_N)\rangle+\zeta_L y_a^{i}\partial_{y_a^{i}}\langle\mathcal{O}(\vec{y}_1,\dots,\vec{y}_N)\rangle+\frac{1}{2}\gamma^{L}_{ij}y_{a}^{j}\partial_{y_a^{i}}\langle\mathcal{O}(\vec{y}_1,\dots,\vec{y}_N)\rangle \nonumber \\ +\frac{1}{8}\left[\gamma^{L}_{ik}\gamma^{L}_{kj}y_a^{j}\partial_{y_a^{i}}+\gamma^{L}_{i\alpha}\gamma^{L}_{j\beta}y_a^{\alpha}y_b^{\beta}\partial_{y_a^{i}}\partial_{y_b^{j}}\right]\langle\mathcal{O}(\vec{y}_1,\dots,\vec{y}_N)\rangle
\end{align}
Next, we average over two long tensor modes and go to Fourier space. After some algebra (and integration by parts), we get,
\begin{equation}
\label{double:soft:tensor}
    \boxed{
\begin{aligned}
\lim_{\vec{q}_1,\vec{q}_2\rightarrow 0}\langle \gamma^{s_1}(\vec{q}_1)\gamma^{s_2}(\vec{q}_2)\mathcal{O}&\rangle'=-\frac{1}{2}\sum_{S}\langle\gamma^{s_1}(\vec{q}_1)\gamma^{s_2}(\vec{q}_2)\gamma^{S}(\vec{q})\rangle'\epsilon^{h}_{ij}(\vec{q})k_{ai}\partial_{k_{aj}}\langle\mathcal{O}(\vec{k}_i,\dots,\vec{k}_N)\rangle'\nonumber \\ &\langle\gamma^{s_1}(\vec{q}_1\gamma^{s_2}(\vec{q}_2)\zeta(-\vec{q}))\rangle' \left[-3(N-1)-k_{ai}\partial_{k_{ai}}\right]\langle\mathcal{O}(\vec{k}_i,\dots,\vec{k}_N)\rangle'
\nonumber\\&+\frac{P(q_1)P(q_2)}{8}\left(\epsilon^{s_1}_{i\alpha}(\vec{q}_1)\epsilon^{s_2}_{\alpha j}(\vec{q}_2)k_{ai}\partial_{k_{aj}}+\epsilon^{s_2}_{i\alpha}(\vec{q}_2)\epsilon^{s_1}_{\alpha j}(\vec{q}_1)k_{ai}\partial_{k_{aj}}\right.\nonumber \\
&\hspace{40pt}\left.+2\epsilon^{s_1}_{i\alpha}(\vec{q}_1)\epsilon^{s_2}_{j\beta}(\vec{q}_2)k_{ai}k_{bj}\frac{\partial^{2}}{\partial k_{a\alpha}\partial_{k_{b\beta}}}\right)\langle\mathcal{O}(\vec{k}_i,\dots,\vec{k}_N)\rangle'
\end{aligned}}
\end{equation}
where $\vec{q}=-\vec{q}_1-\vec{q}_2$. Primes ($\langle\dots\rangle'$) denote that we have stripped of the momentum-conserving delta functions. To the best of our knowledge, this is a novel single-field inflation consistency relation for tensor modes. The first line in the RHS above can come from diagrams given in Figure \ref{all:t:t} with $N=2$. Such diagrams (with internal soft modes) factorise \cite{Joyce:2014aqa,Mirbabayi:2014zpa} and produce 
\begin{align}
\sum_S\langle\gamma^{s_1}(\vec{q}_1)\gamma^{s_2}(\vec{q}_2)\gamma^{S}(\vec{q})\rangle'\frac{1}{P_{\gamma}(q)}\langle\gamma^{S}(-\vec{q})\mathcal{O}(\vec{k}_i,\dots,\vec{k}_N)\rangle'
\end{align}
The second term in the product simplifies further, and we get 
\begin{align}
\frac{1}{2}\sum_{S}\langle\gamma^{s_1}(\vec{q}_1)\gamma^{s_2}(\vec{q}_2)\gamma^{S}(\vec{q})\rangle'\epsilon^{h}_{ij}(\vec{q})k_{ai}\partial_{k_{aj}}\langle\mathcal{O}(\vec{k}_i,\dots,\vec{k}_N)\rangle'
\end{align}
Similarly, the second line in the double-soft formula above can come from diagrams given in Figure \ref{all:t:s} (for $N=2$). In all our soft theorems, we will have such terms which are products of correlators of purely soft modes and some differential operator acting on a hard correlator. These terms arise when three or more soft modes interact at a vertex and then later act as a background for the short modes.   
\section{\texorpdfstring{$N$}{N} soft: scalars}\label{N:soft:scalars}
\begin{figure}[h]
\centering
\begin{minipage}{0.48\textwidth}
\centering
\includegraphics[width=\linewidth]{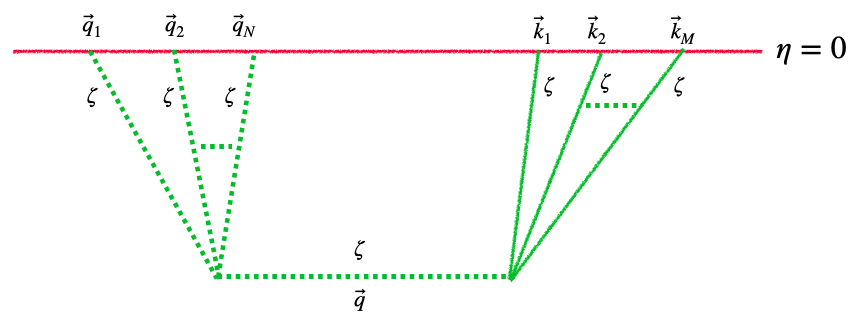}
\caption{$N$ soft scalar modes combine into an internal soft scalar mode which then connects to the hard vertex}\label{all:s:s}
\end{minipage}
\hfill
\begin{minipage}{0.48\textwidth}
\centering
\includegraphics[width=\linewidth]{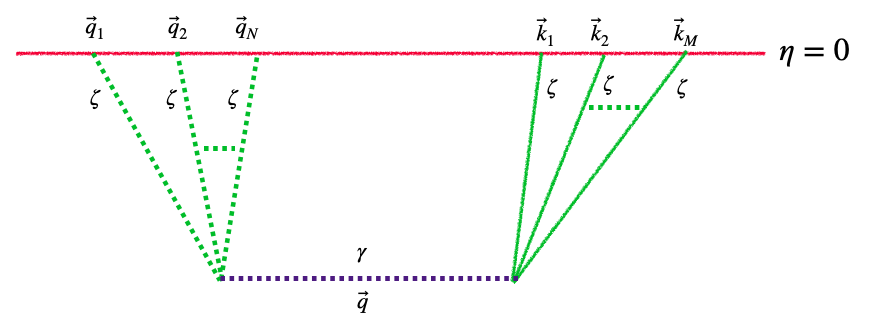}
\caption{$N$ soft scalar modes combine into an internal soft tensor mode which then connects to the hard vertex}
\label{all:s:t}\end{minipage}
\vspace{-1pt}
\end{figure}

\begin{figure}[h]
\centering
\begin{minipage}{0.4\textwidth}
\centering
\includegraphics[width=\linewidth]{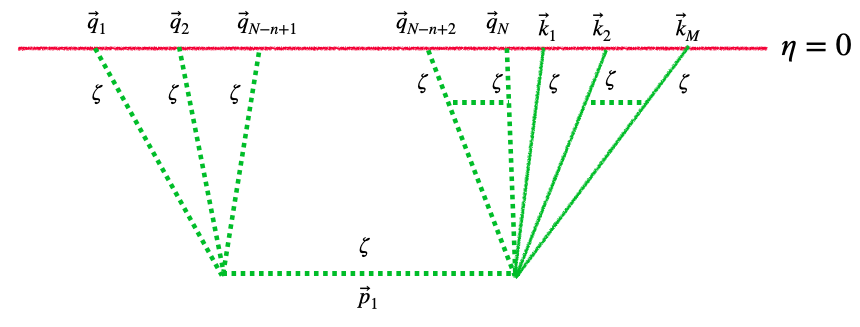}
\caption{$N-n+1$ soft scalar modes combine into an internal soft scalar mode which then connects to the hard vertex. The rest of the soft scalar modes emerge from the hard vertex}
\label{s:s}
\end{minipage}
\hfill
\begin{minipage}{0.48\textwidth}
\centering
\includegraphics[width=\linewidth]{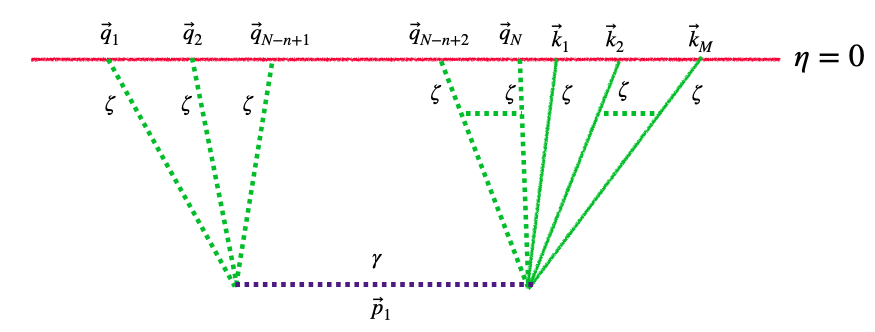}
\caption{$N-n+1$ soft scalar modes combine into an internal soft tensor mode, which then connects to the hard vertex. The rest of the soft scalar modes emerge from the hard vertex}
\label{s:t}
\end{minipage}
\vspace{-20pt}
\end{figure}

The scalar soft theorem with $N$ legs soft was derived in \cite{Joyce:2014aqa} using the 1PI action method. Here, we derive this theorem using the background wave method. Unlike \cite{Joyce:2014aqa}, we also include the graviton exchange contributions where soft scalar modes can combine to produce a soft graviton. Like before, we absorb the long modes in a coordinate rescaling. Since we want $N$ soft modes, we expand $e^{2\zeta_L}$ up to $N^{th}$ power and $(e^{\gamma})_{ij}$ up to linear order (this captures graviton exchange contributions). The transformation that absorbs these reads,
\begin{align}
    \tilde{x}^{i}=x^{i}+\frac{1}{2}x^{j}\gamma_{ij}^{L}\left(1+\zeta_L+\dots+\frac{1}{(N-2)!}\zeta^{N-2}_L\right)+x^{i}\left(\zeta_L+\frac{1}{2!}\zeta^{2}_L+\dots+\frac{1}{N!}\zeta^{N}_L\right)
\end{align}
The background wave method tells us that 
\begin{align}
\langle\zeta_L(\vec{x}_1)\dots\zeta_L(\vec{x}_N)\zeta(\vec{y}_1)\dots\zeta(\vec{y}_M)\rangle=\langle\zeta_L(\vec{x}_1)\dots\zeta_L(\vec{x}_N)\hspace{2pt}\langle\zeta(\vec{\tilde{y}}_1)\dots\zeta(\vec{\tilde{y}}_M)\rangle\rangle
\end{align}
Now, we Taylor expand $\langle\zeta(\vec{\tilde{y}}_1)\dots\zeta(\vec{\tilde{y}}_M)\rangle$, and arrange terms in powers of $\zeta_L$,
\begin{align}\label{Nsoft:taylor}
&\langle\zeta(\vec{y}_1)\dots\zeta(\vec{y}_M)\rangle+\frac{1}{2}\tilde{y}^{i}_{a}\partial_{x^i_{a}}\langle\zeta(\vec{y}_1)\dots\zeta(\vec{y}_M)\rangle+\zeta_Ly^{i}_a\partial_{y^i_{a}} \langle\zeta(\vec{y}_1)\dots\zeta(\vec{y}_M)\nonumber \\&+\sum_{r=1}^{2}\frac{\zeta^{2}_L}{2!}S(2,r)y^{i_1}_{a_1}\dots y^{i_r}_{a_r}\partial_{y^{i_1}_{a_1}}\dots\partial_{y^{i_r}_{a_r}}\langle\zeta(\vec{y}_1)\dots\zeta(\vec{y}_M)\rangle+\dots \nonumber \\
&\dots+\sum_{r=1}^{N}\frac{\zeta^{N}_L}{N!}S(N,r)y^{i_1}_{a_1}\dots y^{i_r}_{a_r}\hspace{2pt}\partial_{y^{i_1}_{a_1}}\dots\partial_{y^{i_r}_{a_r}}\langle\zeta(\vec{y}_1)\dots\zeta(\vec{y}_M)\rangle \nonumber \\
&+\frac{1}{2}\left[\sum_{r=1}^{2}\frac{\zeta_L}{1!}S(2,r)\tilde{y}^{i_1}_{a_1}y^{i_2}_{a_2}\dots y^{i_r}_{a_r}\partial_{y^{i}_{a}}\dots\partial_{y^{i_r}_{a_r}}\langle\zeta(\vec{y}_1)\dots\zeta(\vec{y}_M)\rangle +\dots\right.\nonumber \\
&\left.\dots+\frac{\zeta^{(N-2)}_L}{(N-2)!}\sum_{r=1}^{N-1}S(N-1,r)\tilde{y}^{i_1}_{a_1}y^{i_2}_{a_2}\dots y^{i_r}_{a_r}\dots y^{i_r}_{a_r}\partial_{y^{i}_{a}}\partial_{y^{i_1}_{a_1}}\dots\partial_{y^{i_r}_{a_r}}\langle\zeta(\vec{y}_1)\dots\zeta(\vec{y}_M)\rangle \right]
\end{align}
where $\tilde{y}^{i}_a=\gamma^{L}_{ij}y^{j}_a$ and $S(n,r)=\sum_{i=0}^{r}\frac{(-1)^{r-i}i^{n}}{(r-i)i!}$ is the Stirling number of second kind. Eq. \eqref{Nsoft:taylor} is the generalisation of \eqref{single:soft:taylor}. The final step, as usual, involves multiplying by $N$ soft modes, taking the average, and going to Fourier space. The first term gives only disconnected contributions, so we drop it. The second term is what appears in the single-soft graviton theorem. This, after multiplying with external soft modes and averaging, gives
\begin{align} \label{Eq:all:s:t}
      &\int \prod_{j=1}^{N}d^{3}\vec{q}_j\hspace{1pt}e^{i\sum q^i_{a}x^{i}_a}d^{3}\vec{q}\hspace{1pt}e^{i q^iy^{i}_+} \delta^{(3)}(\sum_{i=1}^{N}\vec{q}_i+\vec{q})\nonumber\\
      &\langle\zeta(\vec{q}_1)\dots\zeta(\vec{q}_N)\gamma^{S}(\vec{q})\rangle'\epsilon^{S}_{ij}\int \prod d^{3}y_b \hspace{1pt}e^{i\sum y^{i}_a k^{i}_a}\delta(\sum \vec{k}_{a})\delta^{\gamma}_{ij}Z
\end{align}
where $\vec{y}_{+}=\frac{\sum_{a=1}^{M}\vec{y}_a}{M}$ and $Z$ is the stripped hard correlator. Similarly, the third term gives,
\begin{align}\label{Eq:all:s:s}
&\int \prod_{j=1}^{N}d^{3}\vec{q}_j\hspace{1pt}e^{i\sum q^i_{a}x^{i}_a}d^{3}\vec{q}\hspace{1pt}e^{i q^iy^{i}_+} \delta^{(3)}(\sum_{i=1}^{N}\vec{q}_i+\vec{q})\nonumber\\
&\langle\zeta(\vec{q}_1)\dots\zeta(\vec{q}_N)\zeta(\vec{q})\rangle'\int \prod d^{3}y_b \hspace{1pt}e^{i\sum y^{i}_a k^{i}_a}\delta(\sum \vec{k}_{a})\delta_{\zeta}Z. 
\end{align}
These two terms can come from diagrams in Figure \ref{all:s:t} \& Figure \ref{all:s:s} (see the end of Sec. \ref{double:soft:gravitons} for a discussion of such terms).
The terms in the second and third lines of \eqref{Nsoft:taylor} are more involved, and we now discuss them in detail. Consider the following  term of this kind $\frac{\zeta^n_L}{n!}\sum_{r=1}^{n\leq N}S(n,r)y^{i_1}_{a_1}\dots y^{i_r}_{a_r}\hspace{2pt}\partial_{y^{i_1}_{a_1}}\dots\partial_{y^{i_r}_{a_r}}\langle\zeta(\vec{y}_1)\dots\zeta(\vec{y}_M)\rangle$. We know that term, \\$\sum_{r=1}^{n}S(n,r)y^{i_1}_{a_1}\dots y^{i_r}_{a_r}\hspace{2pt}\partial_{y^{i_1}_{a_1}}\dots\partial_{y^{i_r}_{a_r}}\langle\zeta(\vec{y}_1)\dots\zeta(\vec{y}_M)\rangle$ in Fourier space (after some integration by parts) reduces to $\int \prod d^{3}y_b \hspace{1pt}e^{i\sum y^{i}_a k^{i}_a}\delta(\sum \vec{k_a})\hspace{1pt} \delta_{\zeta}Z$ for $n=1$ and
$\int \prod d^{3}y_b \hspace{1pt}e^{i\sum y^{i}_a k^{i}_a}\delta(\sum \vec{k_a})\hspace{1pt}\delta^{2}_{\zeta}Z$ for $n=2$ \cite{Joyce:2014aqa}. It is natural to guess that for the general case $n$, the term reduces to $\int \prod d^{3}y_b \hspace{1pt}e^{i\sum y^{i}_a k^{i}_a}\delta(\sum \vec{k_a})\hspace{1pt}\delta^{n}_{\zeta}Z$. We now prove this statement using the method of induction. Assume for the case of $n-1$ soft modes, the following identity holds,
\begin{align},
   \sum_{r=1}^{n-1}S(n-1,r)y^{i_1}_{a_1}\dots y^{i_r}_{a_r}\hspace{2pt}\partial_{y^{i_1}_{a_1}}\dots\partial_{y^{i_r}_{a_r}}\langle\zeta(\vec{y}_1)\dots\zeta(\vec{y}_M)\rangle =\int \prod d^{3}y_b \hspace{1pt}e^{i\sum y^{i}_a k^{i}_a}\delta(\sum \vec{k_a})\hspace{1pt}\delta^{n-1}_{\zeta}Z
\end{align}
Now, pre-multiply both sides by $y^{i_n}_{a_n}\partial_{y^{i_n}_{a_n}}$,
\begin{align}\label{induction}
     \sum_{r=1}^{n-1}S(n-1,r)\left(y^{i_n}_{a_n}\partial_{y^{i_n}_{a_n}}\right)y^{i_1}_{a_1}\dots y^{i_r}_{a_r}\hspace{2pt}\partial_{y^{i_1}_{a_1}}\dots\partial_{y^{i_r}_{a_r}}\langle\zeta(\vec{y}_1)\dots\zeta(\vec{y}_M)\rangle \nonumber \\
     =\left(y^{i_n}_{a_n}\partial_{y^{i_n}_{a_n}}\right)\int \prod d^{3}y_b \hspace{1pt}e^{i\sum y^{i}_a k^{i}_a}\delta(\sum \vec{k_a})\hspace{1pt}\delta^{n-1}_{\zeta}Z
\end{align}
On the RHS of \eqref{induction}, $y^{i_n}_{a_n}\partial_{y^{i_n}_{a_n}}$ can be pulled inside the integral to give 
\begin{align}
\int \prod d^{3}y_b \hspace{1pt}k^{i_n}_{a_n}\partial_{k^{i_n}_{a_n}}\left(e^{i\sum y^{i}_a k^{i}_a}\right)\delta(\sum \vec{k}_{a})\delta^{n-1}_{\zeta}Z
\end{align}
After integration by parts, this simplifies to 
\begin{align}
  \int \prod d^{3}y_b \hspace{1pt}e^{i\sum y^{i}_a k^{i}_a}\delta(\sum \vec{k}_{a})\delta^{n}_{\zeta}Z  
\end{align}
The LHS of \eqref{induction} can be simplified further to get,
\begin{align}\label{proof:LHS1}
    &\sum_{r=1}^{n-1}rS(n-1,r)y^{i_1}_{a_1}\dots y^{i_r}_{a_r}\hspace{2pt}\partial_{y^{i_1}_{a_1}}\dots\partial_{y^{i_r}_{a_r}}\langle\zeta(\vec{y}_1)\dots\zeta(\vec{y}_M)\rangle\nonumber \\
    &+ \sum_{r=2}^{n}S(n-1,r-1)y^{i_1}_{a_1}\dots y^{i_r}_{a_r}\hspace{2pt}\partial_{y^{i_1}_{a_1}}\dots\partial_{y^{i_r}_{a_r}}\langle\zeta(\vec{y}_1)\dots\zeta(\vec{y}_M)\rangle \nonumber\\
   & =\sum_{r=1}^{n}\left[rS(n-1,r)+S(n-1,r-1)\right]y^{i_1}_{a_1}\dots y^{i_r}_{a_r}\hspace{2pt}\partial_{y^{i_1}_{a_1}}\dots\partial_{y^{i_r}_{a_r}}\langle\zeta(\vec{y}_1)\dots\zeta(\vec{y}_M)\rangle
\end{align}
where to rearrange the range of summations, we have used the properties $S(n,0)=S(n,n+1)=0$. The Stirling number of the second kind has a well-known recurrence formula, $S(n,r)=rS(n-1,r)+S(n-1,r-1)$. Using this the LHS of \eqref{induction} simplifies to,
\begin{align}\label{proofLHS2}
\sum_{r=1}^{n}S(n,r)y^{i_1}_{a_1}\dots y^{i_r}_{a_r}\hspace{2pt}\partial_{y^{i_1}_{a_1}}\dots\partial_{y^{i_r}_{a_r}}\langle\zeta(\vec{y}_1)\dots\zeta(\vec{y}_M)\rangle
\end{align}
and we get the following equality,
\begin{align}\label{proof:justscalars}
\sum_{r=1}^{n}S(n,r)y^{i_1}_{a_1}\dots y^{i_r}_{a_r}\hspace{2pt}\partial_{y^{i_1}_{a_1}}\dots\partial_{y^{i_r}_{a_r}}\langle\zeta(\vec{y}_1)\dots\zeta(\vec{y}_M)\rangle \nonumber \\
    =\int \prod d^{3}y_b \hspace{1pt}e^{i\sum y^{i}_a k^{i}_a}\delta(\sum \vec{k}_{a})\delta^{n}_{\zeta}Z  
\end{align}
This completes the proof. Therefore, the term in consideration becomes,
\begin{align}
\frac{\zeta^n_L}{n!}\sum_{r=1}^{n\leq N}S(n,r)y^{i_1}_{a_1}\dots y^{i_r}_{a_r}\hspace{2pt}\partial_{y^{i_1}_{a_1}}\dots\partial_{y^{i_r}_{a_r}}\langle\zeta(\vec{y}_1)\dots\zeta(\vec{y}_M)\rangle=\frac{\zeta^n_L}{n!}\int \prod d^{3}y_b \hspace{1pt}e^{i\sum y^{i}_a k^{i}_a}\delta(\sum \vec{k}_{a})\delta^{n}_{\zeta}Z
\end{align}
Now, when one multiplies this expression by external $N$ long modes and takes the average, we get,
\begin{align}\label{<A>}
    \int\prod_{j=1}^{N}d^{3}\vec{q}_j\hspace{1pt}e^{i\sum q^i_{a}x^{i}_a}\prod_{i=1}^{n}d^{3}\vec{p}_i\hspace{1pt}e^{i\sum p^i_{a}y^{i}_+}\frac{1}{n!}\underbrace{\langle \zeta(\vec{q}_1)\dots\zeta(\vec{q}_N)\zeta(\vec{p}_1)\dots\zeta(\vec{p}_n)\rangle}_{\text{$\langle A \rangle$}}\nonumber\\
    \times \int \prod d^{3}y_b \hspace{1pt}e^{i\sum y^{i}_a k^{i}_a}\delta(\sum \vec{k}_{a})\delta^{n}_{\zeta}Z
\end{align}
where $\vec{y}_{+}=\frac{\sum_{a=1}^{N}\vec{y}_a}{N}$. The connected contribution from $\langle A \rangle$ is relevant only if one considers soft limits of correlators, including loop contributions. In other words, the delta function(s) will not fix all $\vec{p}$\hspace{2pt}'s in terms of external momenta $\vec{q}$\hspace{2pt}'s; there will be left-over momentum integrals over $\vec{p} $\hspace{2pt}'s. However, in this work, we are working at tree-level, and therefore, we only consider contributions where all $\vec{p}$\hspace{2pt}'s can be fixed in terms of external momenta. In the light of this, for $n>2$ (which is the case for the second and third lines of \eqref{Nsoft:taylor}), only disconnected contributions survive. As emphasized in the introduction, we only focus on single exchange contributions. These contributions come from diagrams where two or more external soft modes join at a vertex and are propagated by an internal soft scalar mode. The rest of the external soft modes emerge from the hard vertex (see Figure \ref{s:s}). Such diagrams factorise (due to the presence of internal soft line) \cite{Joyce:2014aqa,Mirbabayi:2014zpa} and therefore, break into the product of connected contributions. This product is what we will get from the disconnected part of $\langle A \rangle$. Let us illustrate this explicitly. We want one internal soft line, which picks one of the $\zeta(\vec{p}_i)$. We are left with $n-1$ $\zeta(\vec{p}_i)$ and each of these we will contract with $n-1$ $\zeta(\vec{q}_i)$. Let us pick these to be $\{\zeta(\vec{q}_{N-n+2}),\dots,\zeta(\vec{q}_{N})\}$. This gives the following product: $(n-1)!\prod_{r=2}^{n} P(q_{N-n+r})\delta^{(3)}(\vec{q}_{N-n+r}+\vec{p}_r)$. The left over $\zeta(\vec{q}_i)$ along with one left over $\zeta(\vec{p}_i)$ gives $n\langle\zeta(\vec{q}_1)\dots\zeta(\vec{q}_{N-n+1})\zeta(\vec{p})\rangle$, where, $-\vec{p}=\sum_{i=1}^{N-n+1}\vec{q}_i$. Therefore, we have the following equation,
\begin{align}
    \langle A \rangle=n!\sum_{\text{perm}}\langle\zeta(\vec{q}_1)\dots\zeta(\vec{q}_{N-n+1})\zeta(\vec{p}_1)\rangle'\left[\prod_{r=2}^{n} P(q_{N-n+r})\delta^{(3)}(\vec{q}_{N-n+r}+\vec{p}_r)\right]\nonumber \\\times\delta^{(3)}(\sum_{i=1}^{N-n+1}\vec{q}_i+\vec{p}_1)
\end{align}
 We now insert the expression above in \eqref{<A>} and get,
 \begin{align}
        &\int \prod_{j=1}^{N}d^{3}\vec{q}_j\hspace{1pt}e^{i\sum q^i_{a}x^{i}_a}\prod_{i=1}^{n}d^{3}\vec{p}_i\hspace{1pt}e^{i\sum p^i_{a}y^{i}_+}\sum_{\text{perm}}\langle\zeta(\vec{q}_1)\dots\zeta(\vec{q}_{N-n+1})\zeta(\vec{p}_1)\rangle' \nonumber \\
        &\times \left[\prod_{r=2}^{n} P(q_{N-n+r})\delta^{(3)}(\vec{q}_{N-n+r}+\vec{p}_r)\right]\delta^{(3)}(\sum_{i=1}^{N-n+1}\vec{q}_i+\vec{p}_1)\int \prod d^{3}y_b \hspace{1pt}e^{i\sum y^{i}_a k^{i}_a}\delta(\sum \vec{k}_{a})\delta^{n}_{\zeta}Z
 \end{align}
 To reduce this formula to familiar form, we change integration variables from $(\vec{p}_1,\dots,\vec{p}_n)$ to $(\vec{p}_1,\dots,\vec{p}_{n-1},\vec{P})$ with $\vec{P}=\sum_{i=1}^{n}\vec{p}_i$ and get,
 \begin{align}
      &\int \prod_{j=1}^{N}d^{3}\vec{q}_j\hspace{1pt}e^{i\sum q^i_{a}x^{i}_a}d^{3}\vec{P}\prod_{i=1}^{n-1}d^{3}\vec{p}_i\hspace{1pt}e^{i P^iy^{i}_+}\sum_{\text{perm}}\langle\zeta(\vec{q}_1)\dots\zeta(\vec{q}_{N-n+1})\zeta(\vec{p}_1)\rangle' \nonumber \\
        &\times \left[\prod_{r=2}^{n-1} P(q_{N-n+r})\delta^{(3)}(\vec{q}_{N-n+r}+\vec{p}_r)\right]P(q_{n})\delta^{(3)}(\vec{q}_{n}+\vec{P}-\sum_{i=1}^{n-1}\vec{p}_i)\nonumber\\
        &\times \delta^{(3)}(\sum_{i=1}^{N-n+1}\vec{q}_i+\vec{p}_1)\int \prod d^{3}y_b \hspace{1pt}e^{i\sum y^{i}_a k^{i}_a}\delta(\sum \vec{k}_{a})\delta^{n}_{\zeta}Z
 \end{align}
 Performing $\vec{p}_1,\dots,\vec{p}_{n-1}$ integrals gives,
 \begin{align} \label{<A1>}
      &\int \prod_{j=1}^{N}d^{3}\vec{q}_j\hspace{1pt}e^{i\sum q^i_{a}x^{i}_a}d^{3}\vec{P}\hspace{1pt}e^{i P^iy^{i}_+}\sum_{\text{perm}}\langle\zeta(\vec{q}_1)\dots\zeta(\vec{q}_{N-n+1})\zeta(\vec{p}_1)\rangle' \left[\prod_{r=2}^{n} P(q_{N-n+r})\right]\nonumber\\
        &\times \delta^{(3)}(\sum_{i=1}^{N}\vec{q}_i+\vec{P})\int \prod d^{3}y_b \hspace{1pt}e^{i\sum y^{i}_a k^{i}_a}\delta(\sum \vec{k}_{a})\delta^{n}_{\zeta}Z
 \end{align}
 where $\vec{p}_1=-\sum_{i=1}^{N-n+1}\vec{q}_i$. In the above formulae, $\sum_{\text{perm}}$ sums all possible permutations of external soft momenta. It is now clear that the structure of these terms (including delta function structures) is the same as those in Eq. \eqref{Eq:all:s:t} \& Eq. \eqref{Eq:all:s:s}. The same structure also appears in \eqref{deltafunction}. Therefore, we can use the method used in Sec. \ref{scalar:single:soft} to reduce everything to a single total momentum conserving delta function (soft+hard momenta both). 
Finally, we now consider terms in the fourth and fifth lines of the Taylor expansion in \eqref{Nsoft:taylor}. Since these terms involve a long graviton mode, they will give rise to contributions where soft external scalar modes combine to give a soft internal graviton mode. Consider the following term of this kind,
 \begin{align}\label{example:term:graviton:exchange}
      &\frac{\zeta^{n-1}_L}{2(n-1)!}\sum_{r=1}^{n\leq N-1}S(n,r)\tilde{y}^{i_1}_{a_1}\dots y^{i_r}_{a_r}\hspace{2pt}\partial_{y^{i_1}_{a_1}}\dots\partial_{y^{i_r}_{a_r}}\langle\zeta(\vec{y}_1)\dots\zeta(\vec{y}_M)\rangle \nonumber \\
    & = \frac{\zeta^{n-1}_L}{2(n-1)!}\sum_{r=1}^{n\leq N-1}S(n,r)\gamma^L_{i_1 j}y^{j}_{a_1}\dots y^{i_r}_{a_r}\hspace{2pt}\partial_{y^{i_1}_{a_1}}\dots\partial_{y^{i_r}_{a_r}}\langle\zeta(\vec{y}_1)\dots\zeta(\vec{y}_M)\rangle
 \end{align}
 Before we average over external soft modes, we will derive the action of the differential operator \eqref{example:term:graviton:exchange} on the stripped hard correlator in Fourier space. We already showed in \eqref{proof:justscalars} that
 \begin{align}
\sum_{r=1}^{n-1}S(n-1,r)y^{i_1}_{a_1}\dots y^{i_r}_{a_r}\hspace{2pt}\partial_{y^{i_1}_{a_1}}\dots\partial_{y^{i_r}_{a_r}}\langle\zeta(\vec{y}_1)\dots\zeta(\vec{y}_M)\rangle \nonumber \\
    =\int \prod d^{3}y_b \hspace{1pt}e^{i\sum y^{i}_a k^{i}_a}\delta(\sum \vec{k}_{a})\delta^{n-1}_{\zeta}Z  
 \end{align}
 where we have just replaced $n$ by $n-1$. Now, pre-multiply the above equation on both sides by $\gamma^L_{i_nj}y^j_{a_n}\partial_{y^{i_n}_{a_n}}$,
 \begin{align}
\left(\gamma^L_{i_nj}y^j_{a_n}\partial_{y^{i_n}_{a_n}}\right)\sum_{r=1}^{n-1}S(n-1,r)y^{i_1}_{a_1}\dots y^{i_r}_{a_r}\hspace{2pt}\partial_{y^{i_1}_{a_1}}\dots\partial_{y^{i_r}_{a_r}}\langle\zeta(\vec{y}_1)\dots\zeta(\vec{y}_M)\rangle \nonumber \\
    =\left(\gamma^L_{i_nj}y^j_{a_n}\partial_{y^{i_n}_{a_n}}\right)\int \prod d^{3}y_b \hspace{1pt}e^{i\sum y^{i}_a k^{i}_a}\delta(\sum \vec{k}_{a})\delta^{n-1}_{\zeta}Z
 \end{align}
 The RHS of the above equation, after some simplifications (integration by parts reads),
 \begin{align}
     \gamma^L_{i_nj}\int \prod d^{3}y_b \hspace{1pt}e^{i\sum y^{i}_a k^{i}_a}\delta(\sum \vec{k}_{a})\left(-k^{i_n}_{a_n}\partial_{k^j_{a_n}}\right)\delta^{n-1}_{\zeta}Z
 \end{align}
 The LHS can also be simplified further by using a procedure identical to \eqref{proof:LHS1}-\eqref{proofLHS2}, and we arrive at the relation, 
 \begin{align}
    & \sum_{r=1}^{n\leq N-1}S(n,r)\gamma^L_{i_1 j}y^{j}_{a_1}\dots y^{i_r}_{a_r}\hspace{2pt}\partial_{y^{i_1}_{a_1}}\dots\partial_{y^{i_r}_{a_r}}\langle\zeta(\vec{y}_1)\dots\zeta(\vec{y}_M)\rangle=\nonumber \\
     &\gamma^L_{i_nj}\int \prod d^{3}y_b \hspace{1pt}e^{i\sum y^{i}_a k^{i}_a}\delta(\sum \vec{k}_{a})\left(-k^{i_n}_{a_n}\partial_{k^j_{a_n}}\right)\delta^{n-1}Z
 \end{align}
 where the LHS above is \eqref{example:term:graviton:exchange} modulo the $\frac{\zeta^{n-1}_L}{2(n-1)!}$ factor. Following steps similar to Eq. \eqref{<A>}-\eqref{<A1>}, we multiply by external soft modes, average, and find,
 \begin{align}
      &\int \prod_{j=1}^{N}d^{3}\vec{q}_j\hspace{1pt}e^{i\sum q^i_{a}x^{i}_a}d^{3}\vec{P}\hspace{1pt}e^{i P^iy^{i}_+}\sum_{\text{perm}}\langle\zeta(\vec{q}_1)\dots\zeta(\vec{q}_{N-n+1})\gamma^{S}(\vec{p}_1)\rangle' \left[\prod_{r=2}^{n} P(q_{N-n+r})\right]\nonumber\\
        &\times \delta^{(3)}(\sum_{i=1}^{N}\vec{q}_i+\vec{P})\epsilon^{S}_{ij}\int \prod d^{3}y_b \hspace{1pt}e^{i\sum y^{i}_a k^{i}_a}\delta(\sum \vec{k}_{a})\delta^{\gamma}_{ij}\delta_{\zeta}^{n-1}Z
 \end{align}
 where $\vec{p}_1=-\sum_{i=1}^{N-n+1}\vec{q}_i$ . In the above formula, there is a sum over helicities, $S$. Such terms (after factorisation due to internal soft mode) can come from diagrams in Figure \ref{s:t}. We now have all the pieces to write the $N$ soft scalar soft theorem. The theorem reads,
 \begin{equation}\boxed{\begin{aligned}
\lim_{\vec{q}_1,\dots,\vec{q}_N\rightarrow0}\langle&\zeta(\vec{q}_1)\dots\zeta(\vec{q}_N)\zeta(\vec{k}_1)\dots\zeta(\vec{k}_M)\rangle'\nonumber \\
&=\langle\zeta(\vec{q}_1)\dots\zeta(\vec{q}_N)\zeta(\vec{q})\rangle\delta_{\zeta}Z-\frac{1}{2}\langle\zeta(\vec{q}_1)\dots\zeta(\vec{q}_N)\gamma^{S}(\vec{q})\rangle\epsilon^{S}_{ij}\delta^{\gamma}_{ij}Z \nonumber \\
&+\sum_{n=2}^{N-1}\sum_{\text{perm}}\langle\zeta(\vec{q}_1)\dots\zeta(\vec{q}_{N-n+1})\zeta(\vec{p}_1)\rangle'\left[\prod_{r=2}^{n} P(q_{N-n+r})\right]\delta^{n}_{\zeta}Z
\nonumber \\
&+\sum_{n=2}^{N-1}\sum_{\text{perm}}\langle\zeta(\vec{q}_1)\dots\zeta(\vec{q}_{N-n+1})\gamma^{S}(\vec{p}_1)\rangle' \epsilon^{S}_{ij}\left[\prod_{r=2}^{n} P(q_{N-n+r})\right]\delta^{\gamma}_{ij}\delta_{\zeta}^{n-1}Z \nonumber \\
&+\left[\prod_{r=1}^{N} P(q_{r})\right]\delta^N_{\zeta} Z
 \end{aligned}}
 \end{equation}
 where the last line come from the special case of \eqref{<A1>}, i.e., for $n=N$. We have isolated this contribution for clarity of presentation. This matches the expression derived in \cite{Joyce:2014aqa}; however, we derived this using the background wave method alone, while that paper derived it using $1$PI action method. Moreover, unlike \cite{Joyce:2014aqa}, the expression we derived includes contributions from soft graviton exchange.
\section{\texorpdfstring{$N$}{N} soft: gravitons}\label{n:soft:gravitons}
\begin{figure}[h]
\centering
\begin{minipage}{0.48\textwidth}
\centering
\includegraphics[width=\linewidth]{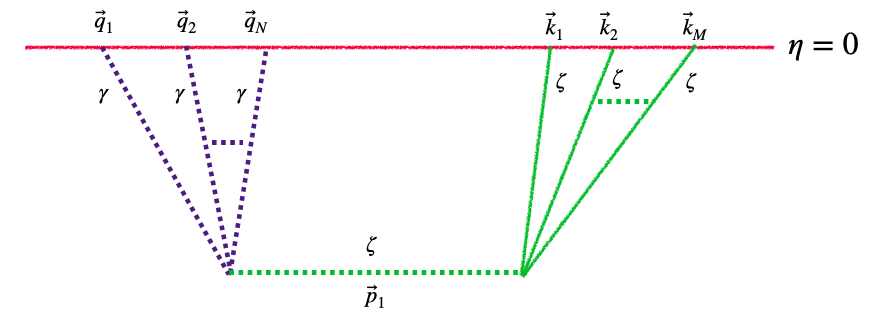}
\caption{$N$ soft tensor modes combine into an internal soft scalar mode which then connects to the hard vertex}
\label{all:t:s}
\end{minipage}
\hfill
\begin{minipage}{0.48\textwidth}
\centering
\includegraphics[width=\linewidth]{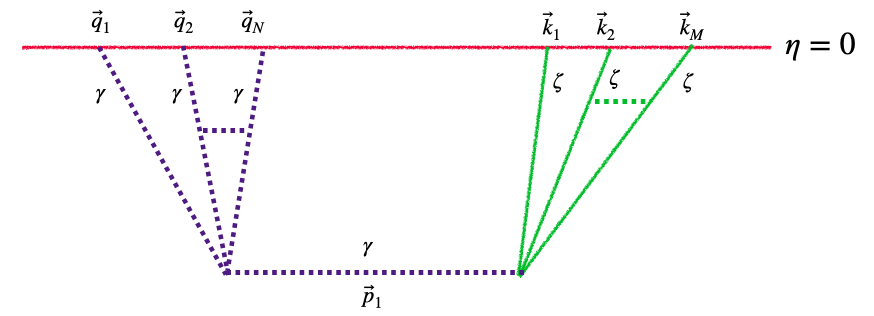}
\caption{$N$ soft tensor modes combine into an internal soft tensor mode which then connects to the hard vertex}
\label{all:t:t}
\end{minipage}
\vspace{-10pt}
\end{figure}

\begin{figure}[h]
\centering
\begin{minipage}{0.48\textwidth}
\centering
\includegraphics[width=\linewidth]{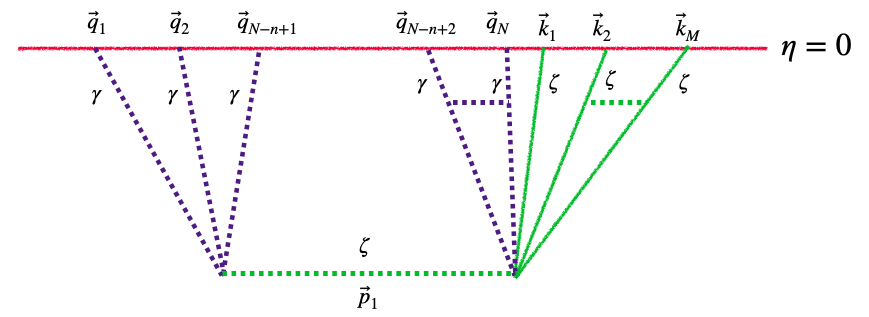}
\caption{$N-n+1$ soft tensor modes combine into an internal soft scalar mode, which then connects to the hard vertex. The rest of the soft tensor modes emerge from the hard vertex}
\label{t:t}
\end{minipage}
\hfill
\begin{minipage}{0.48\textwidth}
\centering
\includegraphics[width=\linewidth]{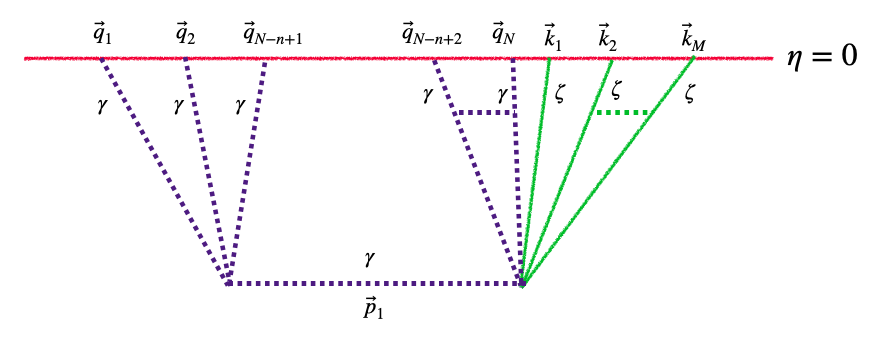}
\caption{$N-n+1$ soft tensor modes combine into an internal soft tensor mode which then connects to the hard vertex. The rest of the soft tensor modes emerge from the hard vertex}
\label{t:s}
\end{minipage}
\end{figure}

In this section, we derive a novel (to the best of our knowledge) $N$ soft theorem for a correlator of the form $\langle \gamma^N\zeta^M\rangle$ using the background wave method. We will follow steps similar to Sec. \ref{N:soft:scalars} and therefore, urge the reader to go through it before reading this section. Here, we expand the graviton long mode (in the metric),$\left(e^{\gamma^L}\right)_{ij}$, up to $N^{th}$ order and the scalar part, $e^{2\zeta_L}$, up to linear order. The latter captures scalar exchange contributions where two or more external soft gravitons combine to form a soft internal scalar mode. Now, as before, we perform a coordinate transformation that absorbs these long modes and gives back the unperturbed metric. This transformation reads,
\begin{align}
\tilde{x}^i=x^i+\zeta_Lx^j\left(I+(\gamma^L/2)+\frac{(\gamma^L/2)^2}{2!}+\dots+\frac{(\gamma^L/2)^{N-2}}{(N-2)!}\right)_{ij}\nonumber \\
+x^j\left((\gamma^L/2)+\frac{(\gamma^L/2)^2}{2!}+\dots+\frac{(\gamma^L/2)^{N}}{N!}\right)_{ij}
\end{align}
Therefore, the desired correlator (using the background wave method) can be obtained as,
\begin{align}
    \langle\gamma^L_{i_1j_1}(\vec{x_1})\dots\gamma^L_{i_N j_N}(\vec{x}_N)\zeta(\vec{y}_1)\dots\zeta(\vec{y}_M)\rangle=\langle\gamma^L_{i_1j_1}(\vec{x_1})\dots\gamma^L_{i_N j_N}(\vec{x}_N)\langle\zeta(\vec{\tilde{y}}_1)\dots\zeta(\vec{\tilde{y}}_M)\rangle\rangle
\end{align}
We now Taylor expand $\langle\zeta(\vec{\tilde{y}}_1)\dots\zeta(\vec{\tilde{y}}_M)\rangle$ and arrange terms in powers of $\gamma_{ij}$ and get,
\begin{align}\label{n:soft:graviton:taylor}
&\langle\zeta(\vec{y}_1)\dots\zeta(\vec{y}_M)\rangle+ \frac{1}{2}\gamma_{ij}y^{j}_a\partial_{y^i_{a}}\langle\zeta(\vec{y}_1)\dots\zeta(\vec{y}_M)\rangle+\zeta_Ly^i_a\partial_{y^i_a}\langle\zeta(\vec{y}_1)\dots\zeta(\vec{y}_M)\rangle
\nonumber \\
&+\sum_{n=2}^{N}\left(\sum_{r=1}^{n}\frac{1}{r!}\sum_{k_1+\dots+k_r=n}\prod_{m=1}^{r}\left[\frac{((\gamma/2)^{k_m})_{i_m j_m}y^{j_m}_{a_m}}{k_m!}\right]\prod_{v=1}^{r}\partial_{y^{i_v}_{a_v}}\langle\zeta(\vec{y}_1)\dots\zeta(\vec{y}_M)\rangle\right) \nonumber \\
&+\zeta_L\sum_{n=2}^{N-1}\left(\sum_{r=1}^{n}\frac{1}{r!}\left[\sum_{k_2+\dots+k_r=n-1}y^{i_1}_{a_1}\prod_{m=2}^{r}\left(\frac{((\gamma/2)^{k_m})_{i_m j_m}y^{j_m}_{a_m}}{k_m!}\right)\right.\right.\nonumber \\
&\left.\left.+\sum_{k_1+\dots+k_r=n-1}\prod_{m=1}^{r}\left(\frac{((\gamma/2)^{k_m})_{i_m j_m}y^{j_m}_{a_m}}{k_m!}\right)\right]\prod_{v=1}^{r}\partial_{y^{i_v}_{a_v}}\langle\zeta(\vec{y}_1)\dots\zeta(\vec{y}_M)\rangle\right)
\end{align}
In the formula above by $((\gamma/2)^{k_m})_{i_m j_m}$, we mean $(i_m, j_m)$ component of $k_m^{th}$ power of graviton matrix. The second and third term in the first line of Taylor expansion (\eqref{n:soft:graviton:taylor}), after averaging over long modes give (modulo the Fourier integral and delta functions), $$\langle\gamma^{s_1}(\vec{q}_1)\dots\gamma^{s_N}(\vec{q}_N)\gamma^{S}(\vec{p}_1)\rangle'\epsilon^S_{ij}(\vec{p}_1)\delta^{\gamma}_{ij}Z \hspace{15pt}\&\hspace{15pt}\langle\gamma^{s_1}(\vec{q}_1)\dots\gamma^{s_N}(\vec{q}_N)\zeta(\vec{p}_1)\rangle'\delta_{\zeta}Z$$ where $Z$ is the stripped hard correlator in Fourier space. These contributions (after factorisation due to internal soft mode) can come from diagrams in Figure \ref{all:t:s} \& \ref{all:t:t}.
Now, consider the second line of the Taylor expansion in \eqref{n:soft:graviton:taylor}. It can be shown (see Appendix \ref{identity:for:n:soft:gravitons}) that,
\begin{align}\label{graviton:identity}
    \sum_{r=1}^{n}\frac{1}{r!}\sum_{k1+\dots+k_r=n}\prod_{m=1}^{r}\left[\frac{((\gamma/2)^{k_m})_{i_m j_m}y^{j_m}_{a_m}}{k_m!}\right]\prod_{v=1}^{r}\partial_{y^{i_v}_{a_v}}\langle\zeta(\vec{y}_1)\dots\zeta(\vec{y}_M)\rangle \nonumber \\
    =\frac{1}{n!}D^{n}_{\gamma}\langle\zeta(\vec{y}_1)\dots\zeta(\vec{y}_M)\rangle
\end{align}
where $D_{\gamma}=(\gamma/2)_{ij}y^{i}_a\partial_{y^{j}_{a}}$. Once we write the hard correlator in terms of its Fourier transform, the RHS becomes,
\begin{align}
    =\left(\prod_{r=1}^{n}\gamma_{i_rj_r}\right)\frac{1}{n!}\int \prod d^{3}y_b \hspace{1pt}e^{i\sum y^{i}_a k^{i}_a}\delta(\sum \vec{k}_{a})\prod_{r=1}^n\delta^{\gamma}_{i_r j_r}Z 
\end{align}
We obtained this formula by repeatedly pushing each $D_\gamma$ inside the Fourier integral and performing integration by parts. Next, we multiple with $N$ external soft gravitons and average,
\begin{align}
     \int \prod_{j=1}^{N}d^{3}\vec{q}_j\hspace{1pt}e^{i\sum q^i_{a}x^{i}_a}&\prod_{i=1}^{n}d^{3}\vec{p}_i\hspace{1pt}e^{i\sum p^i_{a}y^{i}_+}\frac{1}{n!}\sum_{S_1,..,S_n}\langle \gamma^{s_1}(\vec{q}_1)\dots\gamma^{s_N}(\vec{q}_N)\gamma^{S_1}(\vec{p}_1)\dots\gamma^{S_n}(\vec{p}_n)\rangle\nonumber \\
   &  \times \left(\prod_{r=1}^{n}\epsilon^{S_r}_{i_rj_r}(\vec{p}_r)\right)\int \prod d^{3}y_b \hspace{1pt}e^{i\sum y^{i}_a k^{i}_a}\delta(\sum \vec{k}_{a})\prod_{r=1}^n\delta^{\gamma}_{i_r j_r}Z
\end{align}
Following steps similar to Eq. \eqref{<A>}-\eqref{<A1>}, this term reduces to,
\begin{align}
     &\int \prod_{j=1}^{N}d^{3}\vec{q}_j\hspace{1pt}e^{i\sum q^i_{a}x^{i}_a}d^{3}\vec{P}\hspace{1pt}e^{iP^iy^{i}_+}\sum_{\text{perm}}\langle\gamma^{s_1}(\vec{q}_1)\dots\gamma^{s_{N-n+1}}(\vec{q}_{N-n+1})\gamma^{S_1}(\vec{p}_1)\rangle' \epsilon^{S_1}_{i_1 j_1}(\vec{p}_1)\nonumber\\
        &\times\text{sym}\left[\prod_{r=2}^{n} P_{\gamma}(q_{N-n+r})\epsilon^{s_{N-n+r}}_{i_r j_r}(-\vec{q}_{N-n+r})\right]\delta^{(3)}(\sum_{i=1}^{N}\vec{q}_i+\vec{P})\int \prod d^{3}y_b \hspace{1pt}e^{i\sum y^{i}_a k^{i}_a}\delta(\sum \vec{k}_{a})\nonumber \\
        &\hspace{350pt}\times\prod_{r=1}^{n}\delta^{\gamma}_{i_r j_r}Z
\end{align}
where ``$\text{sym}$" denotes the sum over permutations of the helicity labels, keeping the momentum labels fixed, e.g., 
\begin{equation}
\mathrm{sym}\!\left[\epsilon^{s_1}_{ij}(p_1)\,\epsilon^{s_2}_{kl}(p_2)\right]
=
\frac{1}{2!}\left(\epsilon^{s_1}_{ij}(p_1)\,\epsilon^{s_2}_{kl}(p_2)
+
\epsilon^{s_2}_{ij}(p_1)\,\epsilon^{s_1}_{kl}(p_2)\right)\, .
\end{equation}
These contributions can come (after factorisation due to internal soft mode) from diagrams in Figure \ref{t:s}. Next, consider the third and fourth lines of the Taylor expansion in \eqref{n:soft:graviton:taylor}. From Eq. \eqref{graviton:identity}, It can be shown,
\begin{align}
   &\zeta_L\sum_{n=2}^{N-1}\left(\sum_{r=1}^{n}\frac{1}{r!}\left[\sum_{k_2+\dots+k_r=n-1}y^{i_1}_{a_1}\prod_{m=2}^{r}\left(\frac{((\gamma/2)^{k_m})_{i_m j_m}y^{j_m}_{a_m}}{k_m!}\right)\right.\right.\nonumber \\
&\left.\left.+\sum_{k_1+\dots+k_r=n-1}\prod_{m=1}^{r}\left(\frac{((\gamma/2)^{k_m})_{i_m j_m}y^{j_m}_{a_m}}{k_m!}\right)\right]\prod_{v=1}^{r}\partial_{y^{i_v}_{a_v}}\langle\zeta(\vec{y}_1)\dots\zeta(\vec{y}_M)\rangle\right) \nonumber \\
   & =\frac{1}{(n-1)!}\zeta_LD_{\zeta}D^{n-1}_{\gamma}\langle\zeta(\vec{y}_1)\dots\zeta(\vec{y}_M)\rangle
\end{align}
where $D_{\zeta}=y^{i}_a\partial_{y^{i}_a}$. Once again, we write the hard correlator in terms of its Fourier transform, the RHS becomes,
\begin{align}
=\zeta_L\left(\prod_{r=1}^{n-1}\gamma_{i_rj_r}\right)\frac{1}{(n-1)!}\int \prod d^{3}y_b \hspace{1pt}e^{i\sum y^{i}_a k^{i}_a}\delta(\sum \vec{k}_{a})\delta_{\zeta}\prod_{r=1}^{n-1}\delta^{\gamma}_{i_r j_r}Z 
\end{align}
The above relation was derived by repeatedly pushing each operator $D$ inside the Fourier integral and performing integration by parts. Next, we multiple with $N$ external soft gravitons and average,
\begin{align}
    & \int \prod_{j=1}^{N}d^{3}\vec{q}_j\hspace{1pt}e^{i\sum q^i_{a}x^{i}_a}\prod_{i=1}^{n}d^{3}\vec{p}_i\hspace{1pt}e^{i\sum p^i_{a}y^{i}_+}\frac{1}{(n-1)!}\nonumber \\
     &\times \sum_{S_1,..,S_{n-1}}\langle \gamma^{s_1}(\vec{q}_1)\dots\gamma^{s_N}(\vec{q}_N)\zeta_L(\vec{p}_1)\gamma^{S_1}(\vec{p}_2)\dots\gamma^{S_{n-1}}(\vec{p}_n)\rangle
     \left(\prod_{r=1}^{n-1}\epsilon^{S_r}_{i_rj_r}(\vec{p}_{r+1})\right) \nonumber \\
     &\times \int \prod d^{3}y_b \hspace{1pt}e^{i\sum y^{i}_a k^{i}_a}\delta(\sum \vec{k}_{a})\delta_{\zeta}\prod_{r=1}^{n-1}\delta^{\gamma}_{i_r j_r}Z
\end{align}
Following steps similar to Eq. \eqref{<A>}-\eqref{<A1>}, this term reduces to,
\begin{align}
     &\int \prod_{j=1}^{N}d^{3}\vec{q}_j\hspace{1pt}e^{i\sum q^i_{a}x^{i}_a}d^{3}\vec{P}\hspace{1pt}e^{iP^iy^{i}_+}\nonumber\\&
     \times\sum_{\text{perm}}\langle\gamma^{s_1}(\vec{q}_1)\dots\gamma^{s_{N-n+1}}(\vec{q}_{N-n+1})\zeta(\vec{p}_1)\rangle'
       \text{sym}\left[\prod_{r=2}^{n} P_{\gamma}(q_{N-n+r})\epsilon^{s_{N-n+r}}_{i_r j_r}(-\vec{q}_{N-n+r})\right]\nonumber \\
       &\delta^{(3)}(\sum_{i=1}^{N}\vec{q}_i+\vec{P})\int \prod d^{3}y_b \hspace{1pt}e^{i\sum y^{i}_a k^{i}_a}\delta(\sum \vec{k}_{a})\delta_{\zeta}\prod_{r=1}^{n}\delta^{\gamma}_{i_r j_r}Z
\end{align}
These contributions can come (after factorisation due to internal soft mode) from diagrams in Figure \ref{t:t}. We now have all the terms to write the final form of our $N$ soft theorem for gravitons. The theorem reads,
\begin{equation}\boxed{\begin{aligned}
&\lim_{\vec{q}_1,\dots,\vec{q}_N\rightarrow0}\langle\gamma^{s_1}(\vec{q}_1)\dots\gamma^{s_N}(\vec{q}_N)\zeta(\vec{k}_1)\dots\zeta(\vec{k}_M)\rangle'=\langle\gamma^{s_1}(\vec{q}_1)\dots\gamma^{s_N}(\vec{q}_N)\gamma^{S}(\vec{p}_1)\rangle\epsilon^S_{ij}(\vec{p}_1)\delta^{\gamma}_{ij}Z \nonumber \\
&+\langle\gamma^{s_1}(\vec{q}_1)\dots\gamma^{s_N}(\vec{q}_N)\zeta(\vec{p}_1)\rangle\delta_{\zeta}Z \nonumber \\
&+\sum_{n=2}^{N-1}\sum_{\text{perm}}\langle\gamma^{s_1}(\vec{q}_1)\dots\gamma^{s_{N-n+1}}(\vec{q}_{N-n+1})\gamma^{S}(\vec{p}_1)\rangle'\epsilon^{S}_{i_1j_1}\text{sym}\left[\prod_{r=2}^{n} P(q_{N-n+r})\epsilon^{s_{N-n+r}}_{i_r j_r}\right]\prod_{r=1}^{n}\delta^{\gamma}_{i_r j_r}Z
\nonumber \\
&+\sum_{n=2}^{N-1}\sum_{\text{perm}}\langle\gamma^{s_1}(\vec{q}_1)\dots\gamma^{s_{N-n+1}}(\vec{q}_{N-n+1})\zeta(\vec{p}_1)\rangle'\text{sym}\left[\prod_{r=2}^{n} P(q_{N-n+r})\epsilon^{s_{N-n+r}}_{i_r j_r}\right]\delta_{\zeta}\prod_{r=2}^{n}\delta^{\gamma}_{i_r j_r}Z\nonumber \\
&+\text{sym}\left[\prod_{r=1}^{N} P_{\zeta}(q_{r})\epsilon^{s_{r}}_{i_r j_r}\delta^{\gamma}_{i_r j_r}\right]Z
 \end{aligned}}
 \end{equation}
 This reduces to the double-soft graviton theorem derived in Sec. \ref{double:soft:gravitons} for the case of $2$ soft modes, i.e., $N=2$.
 \section{Conclusion and future directions}\label{conclusion}
 In this paper, we derived tree-level multi-soft theorems for gravitons and scalars using the background wave method. Previously, the multi-soft theorem for scalars was derived using the 1PI action method \cite{Joyce:2014aqa}. Here, we derived it using the background wave method and also included the soft graviton exchange contributions (not included previously). The multi-soft theorem we derived for gravitons is a novel relation (to the best of our knowledge). In this case, we also included contributions where a soft scalar mode is exchanged. Some of the future directions include:
 \begin{itemize}
     \item It would be interesting to include loop effects in our multi-soft theorems. The understanding of loop effects in inflation is relatively primitive (see \cite{Tsamis:1996qq,Weinberg:2005vy,Weinberg:2006ac,Senatore:2009cf,Pimentel:2012tw,Lee:2023jby,Bhowmick:2024kld,Bhowmick:2025mxh,Ansari:2025nng} for some developments in this direction).
     \item In this work, we included only single soft exchange contributions in the case of $N>3$ soft theorems. In the future, one would like to incorporate all possible soft exchanges. 
     \item It would be a good cross-check to derive the multi-soft graviton theorem via the 1PI action method. 
     \item In this work, we assumed that both $\zeta$ \& $\gamma$ (in co-moving, transverse-traceless gauge) freeze outside the horizon. It would be interesting to relax this assumption and compute the resulting corrections to our formulae, for instance, in scenarios where an additional shift symmetry is imposed \cite{Finelli:2017fml,Finelli:2018upr}.
   \end{itemize}
 
\appendix
\section{Identity for $N$ Soft: gravitons}\label{identity:for:n:soft:gravitons}
Here, we prove the following identity using induction,
\begin{align}\label{identity}
    \sum_{r=1}^{n}\frac{1}{r!}\sum_{k1+\dots+k_r=n}\prod_{m=1}^{r}\left[\frac{((\gamma/2)^{k_m})_{i_m j_m}y^{j_m}_{a_m}}{k_m!}\right]\prod_{v=1}^{r}\partial_{y^{i_v}_{a_v}}\langle\zeta(\vec{y}_1)\dots\zeta(\vec{y}_M)\rangle \nonumber \\
    =\frac{1}{n!}D^{n}_{\gamma}\langle\zeta(\vec{y}_1)\dots\zeta(\vec{y}_M)\rangle
\end{align}
where $D_{\gamma}=(\gamma/2)_{ij}y^{i}_a\partial_{y^{j}_{a}}$. The base case ($n=1$) holds trivially. Now, let us assume this identity holds for $n$. We will now prove it for $n+1$. Pre-multiply $D_{\gamma}$ on both sides of \eqref{identity}. The RHS becomes $\frac{1}{n!}D^{n+1}_{\gamma}\langle\zeta(\vec{y}_1)\dots\zeta(\vec{y}_M)\rangle$. The LHS breaks into two terms, one where the derivative from $D_{\gamma}$ acts on one of the $r$ $y's$ and the second where the derivative moves past all $y's$ and acts on $\langle\zeta(\vec{y}_1)\dots\zeta(\vec{y}_M)\rangle$. In what follows, we abbreviate $\langle\zeta(\vec{y}_1)\dots\zeta(\vec{y}_M)\rangle$ as $\mathcal{Z}$. Therefore, we have the following LHS,
\begin{align}
    & \sum_{r=1}^{n}\frac{1}{r!}\left[\sum_{k_1+\dots+k_{r+1}=n+1,k_1=1}k_1!\prod_{m=1}^{r+1}\left[\frac{((\gamma/2)^{k_m})_{i_m j_m}y^{j_m}_{a_m}}{k_m!}\right]\prod_{v=1}^{r+1}\partial_{y^{i_v}_{a_v}}\mathcal{Z} \right.\nonumber \\
 &\left.+\sum_{k_1+\dots+k_r=n}r\left[((\gamma/2)^{k_1+1})_{i_1 j_1}((\gamma/2)^{k_2})_{i_2 j_2}...((\gamma/2)^{k_r})_{i_r j_r}\right]\prod_{m=1}^{r}\left[\frac{y^{j_m}_{a_m}}{k_m!}\right]\prod_{v=1}^{r}\partial_{y^{i_v}_{a_v}}\mathcal{Z} \right]
\end{align}
Now, we relabel dummy indices: in the first term above, we replace $r+1\rightarrow r$, and in the second line we replace $k_1+1\rightarrow k_1$. The above expression now reduces to,
\begin{align}\label{identity2}
     & \sum_{r=1}^{n+1}\frac{1}{r!}\left[\sum_{k_1+\dots+k_{r}=n+1,k_1=1}rk_1!\prod_{m=1}^{r}\left[\frac{((\gamma/2)^{k_m})_{i_m j_m}y^{j_m}_{a_m}}{k_m!}\right]\prod_{v=1}^{r}\partial_{y^{i_v}_{a_v}}\mathcal{Z} \right.\nonumber \\
 &\left.+\sum_{k_1+\dots+k_r=n+1,k_1\geq2}rk_1\left[((\gamma/2)^{k_1})_{i_1 j_1}((\gamma/2)^{k_2})_{i_2 j_2}...((\gamma/2)^{k_r})_{i_r j_r}\right]\prod_{m=1}^{r}\left[\frac{y^{j_m}_{a_m}}{k_m!}\right]\prod_{v=1}^{r}\partial_{y^{i_v}_{a_v}}\mathcal{Z} \right] \nonumber \\
& =\sum_{r=1}^{n+1}\frac{1}{r!}\left[\sum_{k_1+\dots+k_{r}=n+1}rk_1\prod_{m=1}^{r}\left[\frac{((\gamma/2)^{k_m})_{i_m j_m}y^{j_m}_{a_m}}{k_m!}\right]\prod_{v=1}^{r}\partial_{y^{i_v}_{a_v}}\mathcal{Z} \right]=\sum_{r=1}^{n+1}\frac{r}{r!}S
\end{align}
where we used the fact that $k_1!$ in the first line of \eqref{identity2} is just $k_1$ since $k_1=1$. We also rearranged the range of summation over $r$ using the fact that terms which don't satisfy the constraint $k_1+...k_r=n+1$ vanish. Since all the $k's$ in the last line of \eqref{identity2} are on equal footing, we can write,
\begin{align}
    rS&=\sum_{k_1+\dots+k_{r}=n+1}(\sum_{b=1}^{r}k_b)\prod_{m=1}^{r}\left[\frac{((\gamma/2)^{k_m})_{i_m j_m}y^{j_m}_{a_m}}{k_m!}\right]\prod_{v=1}^{r}\partial_{y^{i_v}_{a_v}}\mathcal{Z}  \nonumber \\
    &=\sum_{k_1+\dots+k_{r}=n+1}(n+1)\prod_{m=1}^{r}\left[\frac{((\gamma/2)^{k_m})_{i_m j_m}y^{j_m}_{a_m}}{k_m!}\right]\prod_{v=1}^{r}\partial_{y^{i_v}_{a_v}}\mathcal{Z} 
\end{align}
where we used the fact that the sum of $k's$ is equal to $n+1$. This finally gives the LHS as,
\begin{align*}
    \sum_{r=1}^{n+1}\frac{r}{r!}S=\sum_{r=1}^{n+1}\frac{1}{r!}\left[\sum_{k_1+\dots+k_{r}=n+1}(n+1)\prod_{m=1}^{r}\left[\frac{((\gamma/2)^{k_m})_{i_m j_m}y^{j_m}_{a_m}}{k_m!}\right]\prod_{v=1}^{r}\partial_{y^{i_v}_{a_v}}\mathcal{Z} \right]
\end{align*}
and we get our final result,
\begin{align}
    \sum_{r=1}^{n+1}\frac{1}{r!}\left[\sum_{k_1+\dots+k_{r}=n+1}(n+1)\prod_{m=1}^{r}\left[\frac{((\gamma/2)^{k_m})_{i_m j_m}y^{j_m}_{a_m}}{k_m!}\right]\prod_{v=1}^{r}\partial_{y^{i_v}_{a_v}}\mathcal{Z} \right]=\frac{1}{n!}D^{n+1}_{\gamma}\mathcal{Z}\nonumber\\
    or\nonumber \\
    \sum_{r=1}^{n+1}\frac{1}{r!}\left[\sum_{k_1+\dots+k_{r}=n+1}\prod_{m=1}^{r}\left[\frac{((\gamma/2)^{k_m})_{i_m j_m}y^{j_m}_{a_m}}{k_m!}\right]\prod_{v=1}^{r}\partial_{y^{i_v}_{a_v}}\mathcal{Z} \right]=\frac{1}{n!(n+1)}D^{n+1}_{\gamma}\mathcal{Z} \nonumber \\
    =\frac{1}{(n+1)!}D^{n+1}_{\gamma}\mathcal{Z}
\end{align}
This completes the proof.

\acknowledgments
The author acknowledges the support of the Department of Atomic Energy, Government of India, under project no. RTI4019. We also acknowledge support from the National Post-Doctoral Fellowship (N-PDF) provided by Anusandhan National Research Foundation (ANRF). We thank Ashoke Sen for useful discussions throughout this project. We would also like to thank Enrico Pajer and Diptimoy Ghosh for their useful comments and suggestions.




\bibliography{biblio.bib}
\end{document}